\documentclass[aps,prl,twocolumn,nofootinbib,superscriptaddress,showpacs,preprintnumbers,notitlepage]{revtex4-1}

\usepackage{amsmath,amsfonts,amssymb}
\usepackage[utf8]{inputenc}
\usepackage{braket}
\usepackage{graphicx}
\usepackage{bm}
\usepackage{comment}
\usepackage{placeins}
\usepackage{xfrac}
\usepackage[capitalize]{cleveref}
\usepackage{bbold}

\let\revappendix\appendix

\newcommand{\tj}[6]{ \begin{pmatrix}
   #1 & #2 & #3 \\
   #4 & #5 & #6 
  \end{pmatrix}}
\newcommand{\SU}[0]{$\textrm{SU}(2)$}
\newcommand{\h}[1]{\hat{#1}}
\newcommand{\wh}[1]{\widehat{#1}}

\begin{document}
\title{Restricted Boltzmann Machines for Quantum States with Nonabelian or Anyonic Symmetries}

\author{Tom Vieijra}
\affiliation{Department of Physics and Astronomy, Ghent University, B-9000 Ghent, Belgium}

\author{Corneel Casert}
\affiliation{Department of Physics and Astronomy, Ghent University, B-9000 Ghent, Belgium}

\author{Jannes Nys}
\affiliation{Department of Physics and Astronomy, Ghent University, B-9000 Ghent, Belgium}

\author{Wesley De Neve}
\affiliation{Center for Biotech Data Science, Ghent University Global Campus, 21985 Incheon, Republic of Korea}
\affiliation{IDLab, Department of Electronics and Information Systems, Ghent University, B-9052 Ghent, Belgium}

\author{Jutho Haegeman}
\affiliation{Department of Physics and Astronomy, Ghent University, B-9000 Ghent, Belgium}

\author{Jan Ryckebusch}
\affiliation{Department of Physics and Astronomy, Ghent University, B-9000 Ghent, Belgium}

\author{Frank Verstraete}
\affiliation{Department of Physics and Astronomy, Ghent University, B-9000 Ghent, Belgium}

\begin{abstract}
Although artificial neural networks have recently been proven to provide a promising new framework for constructing quantum many-body wave functions, the parameterization of a quantum wavefunction with nonabelian symmetries in terms of a Boltzmann machine inherently leads to biased results due to the basis dependence. We demonstrate that this problem can be overcome by sampling in the basis of irreducible representations instead of spins, for which the corresponding ansatz respects the nonabelian symmetries of the system. We apply our methodology to find the ground states of the one-dimensional antiferromagnetic Heisenberg (AFH) model with spin-\sfrac{1}{2} and spin-1 degrees of freedom, and obtain a substantially higher accuracy than when using the $s_z$-basis as input to the neural network. The proposed ansatz can target excited states, which is illustrated by calculating the energy gap of the AFH model.  We also generalize the framework to the case of anyonic spin chains.
\end{abstract}

\maketitle
\textit{Introduction}---Driven by the rapidly advancing research in artificial intelligence, many-body physics has embraced machine learning (ML) as a powerful tool to tackle non-trivial problems~\cite{carleo_machine_2019}. Applications include the use of neural networks for phase classification~\mbox{\cite{carrasquilla_machine_2017, van_nieuwenburg_learning_2017, morningstar_deep_2018, beach_machine_2018, chng_machine_2017, venderley_machine_2018, casert_interpretable_2019, liu_discriminative_2018, zhang_machine_2018}}, accelerating Monte Carlo algorithms~\mbox{\cite{wang_exploring_2017, liu_self-learning_2017, liu_self-learning_2017-1, huang_accelerated_2017, bojesen_policy-guided_2018}}, and ML-based generative modeling of distributions in many-body physics~\mbox{\cite{torlai_learning_2016, liu_simulating_2017,mills_phase_2017, carleo_solving_2017}}. The connection between the renormalization group and (deep) learning has also been highlighted~\mbox{\cite{koch-janusz_mutual_2018, li_neural_2018, mehta_exact_2014, iso_scale-invariant_2018}}. 

Quantum mechanical spin systems play a key role in the field of many-body physics. In Ref.~\cite{carleo_solving_2017}, a particular class of artificial neural networks (ANN), namely restricted Boltzmann machines (RBM), is introduced as a variational ansatz for wave functions of many-body spin systems. The versatility of the ANN ansatz has been illustrated in studies of bosonic systems~\cite{saito_solving_2017, saito_machine_2017}, (chiral) topological states~\cite{glasser_neural-network_2018,deng_machine_2017,zheng_restricted_2018}, frustrated systems~\cite{liang_solving_2018, choo_study_2019} and open systems \cite{hartmann_neural-network_2019, yoshioka_constructing_2019, nagy_variational_2019, vicentini_variational_2019}. The RBM ansatz has been studied from the perspective of entanglement~\cite{deng_quantum_2017}, and was shown to embody volume-law entanglement. In this light, the connections and differences between RBMs and matrix product states have been laid out~\cite{chen_equivalence_2018, pastori_generalized_2019}.

The invariance of wave functions under symmetries, in particular \SU~symmetry, is important for applications such as quantum chemistry~\cite{szabo_modern_1996,gunst_three-legged_2019} and the description of spin liquids~\cite{zhou_quantum_2017,savary_quantum_2017}. In quantum chemistry, wave functions are eigenfunctions of the total angular momentum operators $\wh{J}^2$ and $\wh{J}_z$, and spin liquid states do not break any symmetry of the Hamiltonian. The capacity of RBMs to capture long-range correlations and their independence of the problem's geometry make them a prime candidate for these applications.

In this Letter, we introduce a methodology for constructing an RBM variational wave function that transforms as an irreducible representation of~\SU. This wave function is designed to have a well-defined total angular momentum. Hence the method provides direct access to the construction of excited states. The challenge of imposing physical symmetries on the ANN ansatz has been addressed for finite abelian symmetry groups~\cite{choo_symmetries_2018}, but this approach is not directly applicable to nonabelian symmetries. The proposed ansatz is not restricted to \SU~symmetry  and  can  be  applied  to  other nonabelian symmetries, as well as anyonic spin chains.

\textit{RBM wave functions}---We use RBMs, which are energy-based generative neural networks, as a variational ansatz for quantum many-body spin systems. In the context of this work, the RBM models a distribution $\Psi(x_1,...,x_{N}) \equiv \Psi(\mathbf{x})$ of the variables $x_{i\in \{1,...,N \}}$, characterized by the energy function
\begin{equation}
    E_{RBM}(\mathbf{x}, \mathbf{h} ; \mathcal{W}) = \sum_i^{N}\sum_j^{M} w_{ij} x_i h_j + \sum_{i}^{N} a_i x_i + \sum_{j}^{M} b_j h_j\;.
    \label{eq:RBME}
\end{equation}
Here, $h_{j\in \{1,...,M\}} \in \{-1,1\}$ is a set of binary latent variables. The set $\mathcal{W} = \{w_{ij},a_i,b_j\}$ are variational parameters: $w_{ij}$ are the weights connecting variables $x_i$ and $h_j$, and $a_i$~($b_j$) are the biases of the physical (latent) variables $x_i$~($h_j$). The ratio of the number of latent variables $M$ and the number of physical variables $N$, defined by $\alpha \equiv M/N$, is a measure of the complexity of the model. 

The RBM was introduced in Ref.~\cite{carleo_solving_2017} as a variational ansatz for quantum many-body wave functions by modeling the probability amplitudes $\Psi(\mathbf{x})$ of the wave function \mbox{$\ket{\Psi} = \sum_{\mathbf{x}} \Psi(\mathbf{x}) \ket{\mathbf{x}}$} as the marginalized Boltzmann distribution
\begin{equation}
    \Psi(\mathbf{x};\mathcal{W}) \equiv \sum_{\mathbf{h}} e^{-E_{RBM}(\mathbf{x}, \mathbf{h} ; \mathcal{W})} \;,\quad\mathcal{W}\subset\mathbb{C}.
    \label{eq:psiRBM}
\end{equation}
This model is an extension of the RBM that is used to represent classical probability distributions as a marginalization over hidden variables of a Boltzmann  distribution. In this work, $x_i$ is not necessarily a binary variable but remains discrete. We therefore replace $w_{ij} x_i$ with a set of general discrete functions $w_{ij}[x_i]$.  This has the effect of transforming the variables $x_i$ non-linearly before constructing the RBM energy function of Eq.~\eqref{eq:RBME}.

To find eigenstates of a Hamiltonian $\wh{H}$, we use the variational principle to minimize the energy functional
\begin{equation}
    E(\Psi;\mathcal{W}) = \frac{\braket{\Psi|\wh{H}|\Psi}}{\braket{\Psi|\Psi}},
    \label{eq:e_func}
\end{equation}
with respect to the parameters $\mathcal{W}$, for which we use the stochastic reconfiguration method~\cite{sorella_green_1998}.
\begin{figure}[t]
\includegraphics{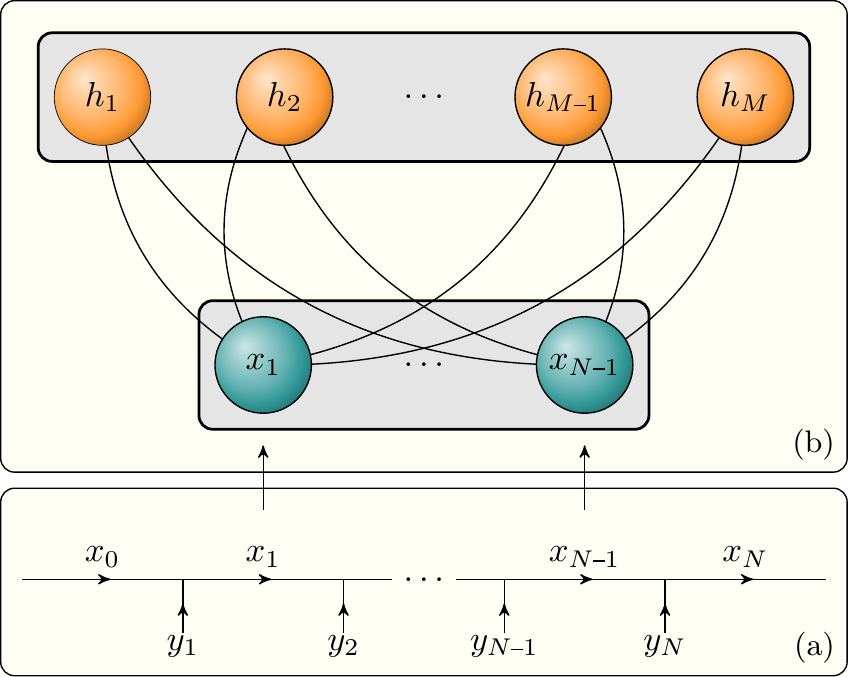}
\caption{Graphical representation of the variational wave function proposed in this work. (a) A basis in which all the degrees of freedom $y_{i \in \{1,...,N\}}$ are coupled is proposed. The resulting intermediate degrees of freedom $x_{i \in \{1,...,N-1\}}$ act as the input of the RBM, as represented in (b). The RBM defines the expansion coefficients of the wave function in this basis. The optimal coefficients minimize the energy functional of Eq.~\eqref{eq:e_func}.}
  \label{fig:fusionRBM}
\end{figure} 

\textit{\SU~symmetry}---Hamiltonians with spin-rotation symmetry are omnipresent throughout the various domains of quantum many-body physics. Examples of \SU-symmetric quantum-mechanical spin chains are the antiferromagnetic Heisenberg model (AFH), the bilinear-biquadratic model, and the Majumdar-Ghosh model. In order to illustrate the potential of the proposed methodology, we focus on the AFH Hamiltonian
\begin{equation}
    \wh{H}_{AFH} = \sum_{i=1}^{N-1} \h{\mathbf{s}}_i \cdot \h{\mathbf{s}}_{i+1} \; ,
\end{equation}
with spin-\sfrac{1}{2} and spin-1 degrees of freedom in one dimension, which we study with open boundary conditions.

The main contribution of this work is the construction of a method to include \SU~symmetry in ANN wave functions.  This procedure yields wave functions $\ket{J\,M_J}$ with well-defined total angular momentum $J$ and projection $M_J$.  These wave functions belong to the subspace spanned by states with quantum numbers $J$ and $M_J$ of the full Hilbert space of the system. Part of the energy spectrum of the system can be uncovered by finding the variational minima in the subspaces with fixed quantum numbers $\ket{J\,M_J}$. For example, the gap of the AFH can be calculated after constructing the ground state $\ket{J=0\,M_J=0}$ and the first excited state $\ket{J=1\,M_J \in \{-1,0,1\}}$. 

To construct wave functions $\ket{J\,M_J}$, we use Clebsch-Gordan coefficients to couple two spins, $\h{\mathbf{s}}_1$ and $\h{\mathbf{s}}_2$, to a single angular momentum degree of freedom $\h{\mathbf{j}}_2 = \h{\mathbf{s}}_1 + \h{\mathbf{s}}_2$
\begin{equation}\begin{split}
    \ket{s_1\,s_2;j_2\, m_{j_2}} = \sum_{m_{s_1},m_{s_2}} &\braket{s_1\,m_{s_1},s_2\,m_{s_2} | j_2\,m_{j_2}}\\ &\ket{s_1\,m_{s_1},s_2\,m_{s_2}}.
    \end{split}
    \label{eq:CG}
\end{equation}
Given a system of $N$ spins $\h{\mathbf{s}}_{i \in \{1,...,N\}}$, we construct states with total angular momentum $J$ by starting at the left side of the chain and using Eq.~\eqref{eq:CG} to couple the first two spins to the angular momentum $j_2$. Next, $j_{2}$ is coupled to $s_{3}$.  This process is repeated till reaching the end of the chain, resulting in a total angular momentum $J$. This is depicted in Fig.~1(a) with $y_i \equiv s_i$, $x_i \equiv j_i$ and $x_0 = 0$, $x_N = J$.
Given the large amount of intermediate couplings, it is non-trivial and numerically challenging to transform the wave function in the coupled basis back into the $\h{\mathbf{s}}_{i \in \{1,...,N\}}$ basis. As will become clear, however, observables of the studied quantum systems can be reliably and efficiently computed in the basis of coupled angular momenta.
In the ansatz proposed in Ref.~\cite{carleo_solving_2017}, the spin projections $m_{s_i}$ of $\h{\mathbf{s}}_i$ are used as input values for the RBM in Eq.~\eqref{eq:psiRBM}. Rather, we use the intermediate degrees of freedom $j_k$ as input (Fig.~\ref{fig:fusionRBM}(b)), which produces the wave function 
\begin{equation}
    \ket{\Psi} = \sum_{j_k} \Psi (j_1,...,j_{N-1}) \ket{j_1,...,j_{N-1};J\,M_J},
    \label{eq:ansatz}
\end{equation} 
where $\sum_{j_k}$ denotes a summation over all physically allowed configurations $j_{k \in \{1,...,N-1\}}$. Eq.~\eqref{eq:ansatz} transforms as an irreducible representation of \SU, labeled by total angular momentum $J$, with dimension $2J+1$.  For the states with $J=0$ (of which the ground state of the \mbox{spin-\sfrac{1}{2}} AFH is an example), the state is manifestly invariant under \SU~transformations, as the irreducible representation has dimension $1$.  More information on this basis transformation can be found in the Supplemental Material (SM)~\cite{SM}.  The above procedure can be readily extended to other nonabelian symmetries by decomposing the degrees of freedom in irreducible representations of the  symmetry  group  and  finding  the  equivalent  of  the Clebsch-Gordan coefficients in Eq.~\eqref{eq:CG} to relate the irreducible representation of a system to those of its subsystems.

\begin{figure}
\includegraphics{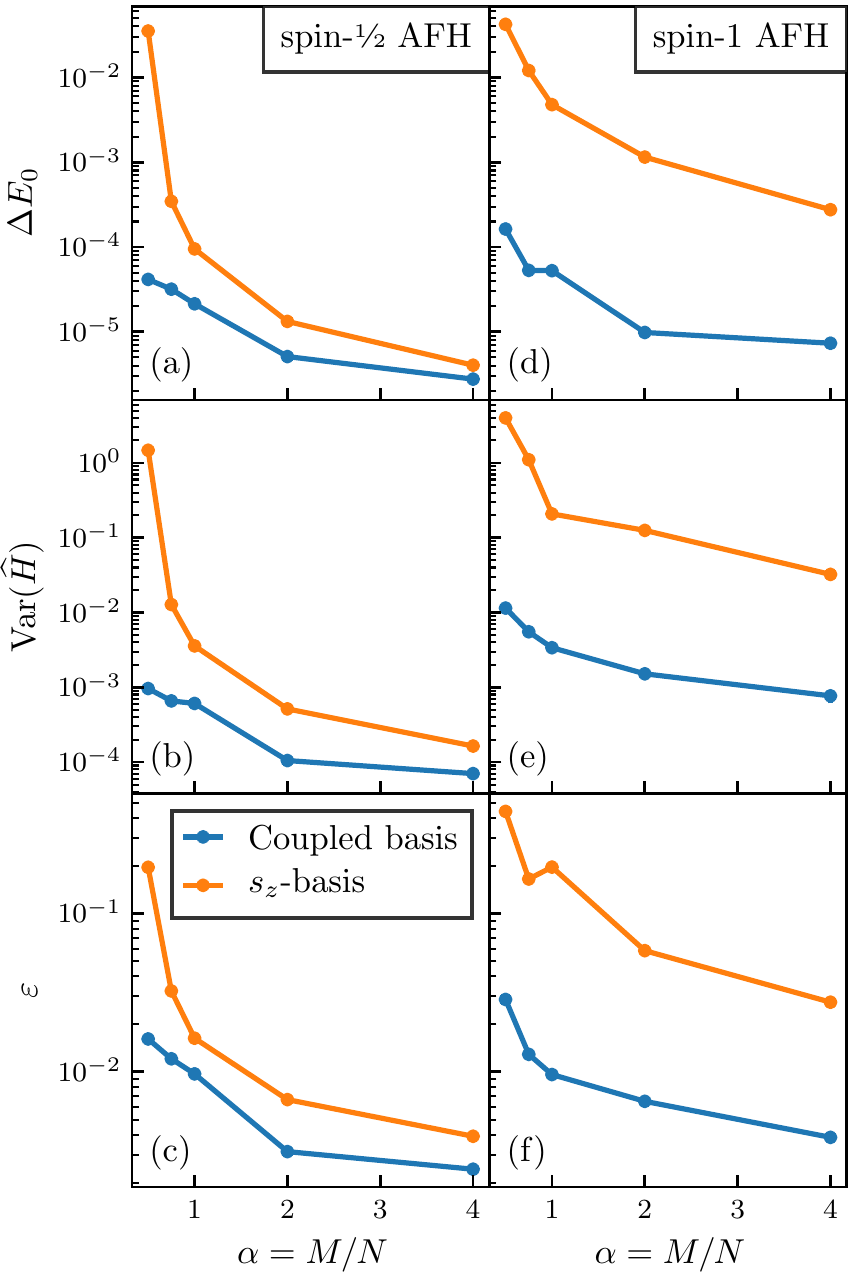}
\caption{Convergence of the ground-state properties as a function of the ratio $\alpha$ of hidden to visible units, for the spin-\sfrac{1}{2} (left column, system size $L=22$), and spin-1 (right column, $L=12$) AFH. Top panels: energy error relative to the exact ground state energy; Middle panels: variance of the Hamiltonian; Bottom panels: content $\varepsilon$ of excited states in the ground-state approximation.  The used hyperparameters can be found in the SM~\cite{SM}.} 
  \label{fig:both_quality}
\end{figure} 
\begin{figure*}[t]
\includegraphics{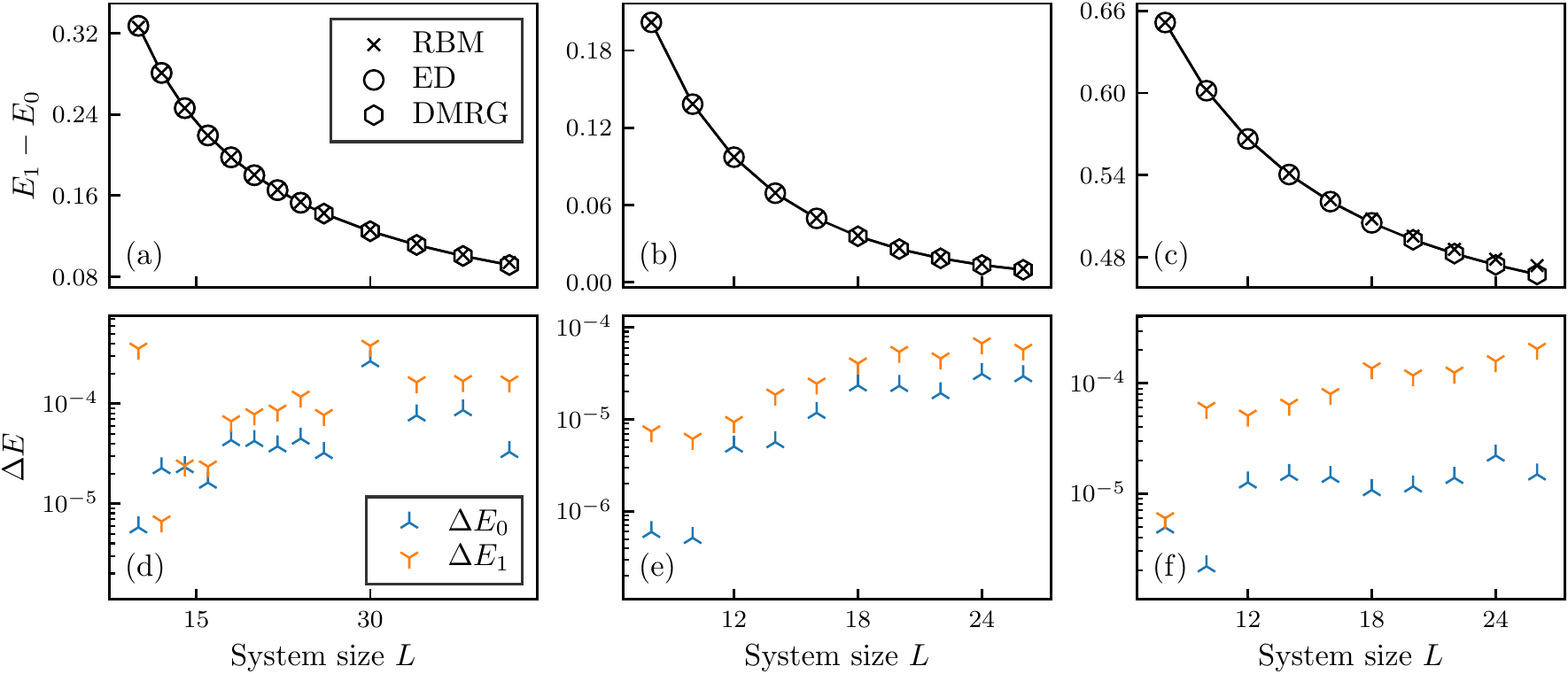}
\caption{(a-c) Energy gaps $G=E_1-E_0$ between the first excited state and the ground state for different system sizes. (d-f) Relative energy errors for the ground states and the first excited states.  The variational energies are compared to those of ED or DMRG. (a, d) Spin-\sfrac{1}{2} AFH ($\alpha=1$), (b, e) spin-1 AFH ($\alpha=2$) and (c, f) spin-1 AFH with spin-\sfrac{1}{2} edges ($\alpha=2$).  The energies of the DMRG simulations were converged to sufficiently high precision without exploiting \SU~symmetry.}
  \label{fig:gaps}
\end{figure*} 
\textit{Spin-\sfrac{1}{2} AFH}---Finding the ground state of the \mbox{spin-\sfrac{1}{2}} AFH in the coupled spin basis amounts to variationally minimizing the energy functional of Eq.~\eqref{eq:e_func} in the subspace defined by total angular momentum $J=0$. The accuracy of the variational wave function can be assessed by comparison to energies obtained with exact diagonalization (ED) \mbox{$\Delta E_0 = |(E_0 - E_{0,exact})/E_{0,exact}|      $}, the magnitude of the variance of the Hamiltonian \mbox{$\textrm{Var}(\wh{H}) = \langle \wh{H}^2 \rangle - \langle \wh{H} \rangle^2$}, and the weight $\varepsilon$ of excited states in the wave function. The latter can be found by writing \mbox{$\ket{\Psi} = \sqrt{1-\varepsilon^2} \ket{\Psi_{GS}} + \varepsilon \ket{\Psi_{\perp}}$}, where $\ket{\Psi_{GS}}$ is the exact ground state wave function and $\ket{\Psi_{\perp}}$ is a normalized superposition of states perpendicular to $\ket{\Psi_{GS}}$. The parameter $\varepsilon$ is a measure for the accuracy of the variational ground state as it measures the spurious content in $\ket{\Psi}$. An upper bound on $\varepsilon$ is given by the relation $\textrm{Var}(\wh{H}) \geq \varepsilon^2 G^2$, with $G\equiv E_1-E_0$ the difference between the energies of the first excited state and the ground state.

In Fig.~\ref{fig:both_quality}(a-c), we compare these convergence criteria for the ground states obtained with the RBM ansatz in the coupled basis and in the $s_z$-basis, as a function of the ratio $\alpha$. Both the relative energy error and $\textrm{Var}(\wh{H})$ are systematically lower when using the coupled basis compared to the $s_z$-basis. For small systems, where ED is feasible, the parameter $\varepsilon$ can be determined exactly by computing the overlap $\braket{\Psi_{GS}|\Psi} = \sqrt{1-\varepsilon}$. We obtain consistently lower values of $\varepsilon$ in the coupled basis.  Related to this, we see that the spin-spin correlators are consistently better described, and inherently unbiased, in the coupled basis.  This is described in the SM~\cite{SM}.

Also with the eye on gaining profound insight in the structure of the wave function, our methodology offers opportunities by studying the weight of the expansion coefficients in the ansatz of Eq.~\eqref{eq:ansatz}. We find that the basis state with the largest modulus has all pairs of neighboring spins coupled to a singlet. Next in importance are states with two neighboring triplets coupled to a singlet, on a background of singlets. More information on the structure of the wave function can be found in the SM~\cite{SM}.

The introduction of the coupled basis allows us to find the variational minimum of the energy functional of Eq.~\eqref{eq:e_func} in a subspace with specific $\ket{J\, M_J}$, which enables us to construct excited states. We demonstrate this by calculating the energy difference between the lowest lying eigenstate in the subspaces defined by $\ket{J=1 \, M_J=0}$ and $\ket{J=0\, M_J=0}$. As the AFH is critical, the gap $G$ vanishes as $G \propto N^{-1}$ for $N \rightarrow \infty$. The gap as a function of system size is depicted in Fig.~\ref{fig:gaps}(a) and matches results obtained with ED or density matrix renormalization group (DMRG). The relative energy errors $\Delta E_0$ on the ground state range from $\mathcal{O}(10^{-5})$ for the smallest system sizes to $\mathcal{O}(10^{-4})$ for larger systems. The errors on the excited state energies  $\Delta E_1$ are generally slightly larger.

\textit{Spin-1 AFH}---Physically, the spin-1 AFH is inherently different from the spin-\sfrac{1}{2} AFH.  Whereas the spin-\sfrac{1}{2} AFH is gapless in the thermodynamic limit, the spin-1 AFH has a fourfold degenerate ground state (consisting of a spin singlet and a spin triplet), above which a gap exists.  The degeneracy of the ground state arises from the presence of effective spin-\sfrac{1}{2} degrees of freedom at the edges of the system.  The interaction between these effective degrees of freedom is exponentially suppressed with system size, resulting in two free spin-\sfrac{1}{2} degrees of freedom in the thermodynamic limit~\cite{white_numerical_1993}. For finite systems, the ground state is non-degenerate and a spin singlet, in accordance with Marshall's theorem~\cite{marshall_antiferromagnetism_1955}. The physical differences between the \mbox{spin-\sfrac{1}{2}} and spin-1 AFH make it interesting to investigate the representational ability of RBMs in both cases.  Figs.~\ref{fig:both_quality}(d-f) shows $\Delta E$, $\textrm{Var}(\wh{H})$ and $\varepsilon$ as a function of the ratio $\alpha$ in the $s_z$-basis and the coupled basis. Across the whole range of $\alpha \in [0.5,4]$, the level of accuracy quantified by these measures is improved by at least an order of magnitude in the coupled basis as compared to the $s_z$-basis. These results are indicative for the effectiveness of the coupled basis. The structure of the wave function is similar to that of the spin-\sfrac{1}{2} Heisenberg model, and is described in detail in the SM~\cite{SM}. 

Figure~\ref{fig:gaps}(b) shows the energy gap $G$ between the ground state $\ket{J=0\,M_J=0}$ and the first excited state $\ket{J=1\,M_J=0}$. For the energy gap, excellent agreement is reached between the RBM and ED/DMRG methods for different system sizes. The relative energy error on the ground state is $\mathcal{O}(10^{-6})$ for the smallest system sizes, and settles to $\mathcal{O}(10^{-5})$ for larger system sizes, while that of the excited state is $\mathcal{O}(10^{-5})$ to $\mathcal{O}(10^{-4})$.  In Fig.~\ref{fig:gaps}(c), the energy gap of the spin-1 AFH with physical spin-\sfrac{1}{2} degrees of freedom on the edges is shown.  The introduction of spin-\sfrac{1}{2} edges lifts the degeneracy of the ground state, and introduces a gap in the system, corresponding to the Haldane gap in the thermodynamic limit. Fig.~\ref{fig:gaps}(c) shows that the RBM ansatz in the coupled basis can represent gapped systems accurately.

\begin{figure}[b!]
\includegraphics{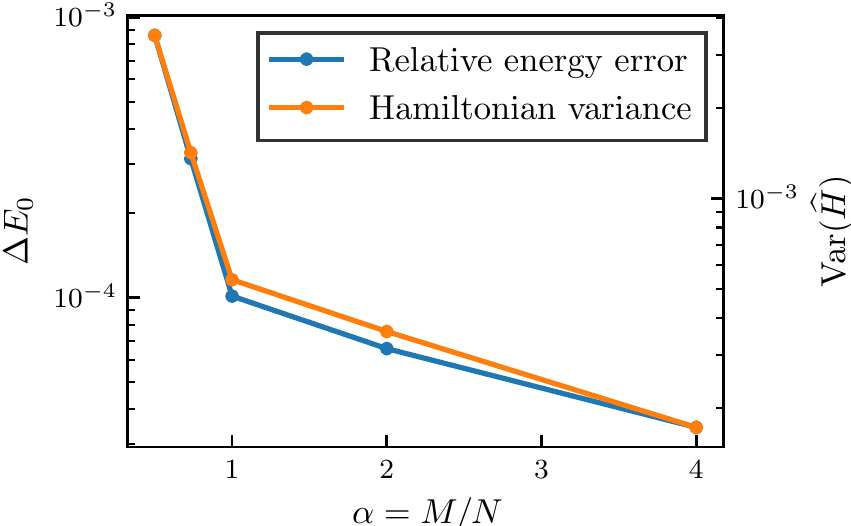}
\caption{Relative energy error $\Delta E_0$ and variance of the Hamiltonian $\textrm{Var}(\wh{H}_{GC})$ of the ground state of the $L=30$ anyonic golden chain.  Both measures decrease with the ratio $\alpha$.}
  \label{fig:anyon}
\end{figure}
\textit{Anyonic golden chain}---As a novel application of the RBM ansatz, we turn to the simulation of anyonic systems. Anyons are defined as degrees of freedom that do not obey the mutual statistics of fermions or bosons~\cite{trebst_short_2008}. They arise in the fractional quantum Hall effect~\cite{arovas_fractional_1984} and play a role in quantum computation~\cite{kitaev_fault-tolerant_2003,nayak_non-abelian_2008}. Here, we study Fibonacci anyons, defined by one anyon type $\tau$ along with the trivial vacuum state $\mathbb{1}$.  Anyonic many-particle systems are most easily described by fusing the different anyons. For Fibonacci anyons, the fusion rules are given by $\mathbb{1} \otimes \tau = \tau \otimes \mathbb{1} = \tau$ and $\tau \otimes \tau = \mathbb{1} \oplus \tau$. For a one-dimensional system, fusing the $N$ anyons can be done in a linear fashion from left to right 
(Fig.~\ref{fig:fusionRBM}(a)), where $y_{i\in\{1,...,N\}} \equiv \tau$ and $x_{i\in\{1,...,N-1\}} \equiv \mathbb{1},\tau$. This construction is called a fusion tree.  We choose the boundary conditions $x_0 = x_N = \mathbb{1}$. Interacting anyons can be described on the level of the fused anyons. An example is the golden chain, which is reminiscent of the AFH model, defined by the Hamiltonian 
\begin{equation}
\wh{H}_{GC} = -\sum_{i=1}^{N-1} \wh{\mathcal{P}}^{\mathbb{1}}_{i,i+1},
\end{equation}
where $\wh{\mathcal{P}}^{\mathbb{1}}_{i,i+1}$ is the projector on the vacuum fusion channel of the anyons with indices $i$ and $i+1$.  The model defined by $\wh{H}_{GC}$ is critical.

We exploit RBMs as a variational ansatz to find the ground state of $\wh{H}_{GC}$. The results of the intermediate fusings are used as input for the RBM (Fig.~\ref{fig:fusionRBM}(b)).  The relative error on the ground-state energy $\Delta E_0$, and the variance of the Hamiltonian $\textrm{Var}(\wh{H}_{GC})$ are shown in Fig.~\ref{fig:anyon} for different values of $\alpha = M/N$.  The relative energy error is below $\mathcal{O}(10^{-3})$ for all values of $\alpha$, reaching $\mathcal{O}(10^{-5})$ for $\alpha=4$.  The variance of the Hamiltonian follows the same trend, ranging from $\mathcal{O}(10^{-3})$ to $\mathcal{O}(10^{-4})$.

\textit{Conclusion}---In this paper, we have extended the variational class of artificial neural network states for spin systems to include nonabelian symmetries, in particular the \SU~spin-rotation symmetry. Hereto, we have formulated a restricted Boltzmann machine (RBM) ansatz using configurations labeled by intermediate spin coupling quantum numbers. Our numerical findings provide convincing support for using these quantities, which do not depend on the choice of local basis for the spins, as input variables to the RBM. Indeed, our ansatz obtains a higher precision in the ground-state energy for the same amount of variational freedom, compared to previous studies using the basis-dependent local spin projections as input variables. In doing so, the states we construct have well-defined total angular momentum $J$ and projection $M_J$ quantum numbers and can also be used to target the lowest-lying excited states of the system. Another application which makes use of the specific structure exploited by our ansatz are anyonic chains, for which our ansatz also accurately captures the ground state.

The approach presented here can be used to model spin liquid states, which are invariant under symmetries of the Hamiltonian. Likewise, our method could be adopted for the determination of wave functions in a quantum chemistry context.  Both these applications are particularly suitable to be studied with RBMs, due to their potential to model long-range entanglement and correlations, and due to the geometrically independent way of modelling correlations in RBMs.  Furthermore, our approach can be generalized for systems in more than one dimension, by defining an appropriate coupling scheme.  These include coupling in the form of a graph, such as a tensor network \cite{singh_tensor_2012}, or coupling in a tree-like fashion.  We stress that our approach is independent of the form of the variational state and is applicable to other variational wave functions.

\begin{acknowledgements}
We thank Stijn De Baerdemacker for useful discussions. We also thank the anonymous referee for suggesting to study correlation functions. The software for this work was built on the \textsc{NetKet} library~\cite{carleo_netket:_2019}. The DMRG results were obtained with the \textsc{ALPS} library \cite{bauer_alps_2011}. The computational resources (Stevin Supercomputer Infrastructure) and services used in this work were provided by the VSC (Flemish Supercomputer Center), and the Flemish Government -- department EWI. This work was supported by Ghent University, Research Foundation Flanders (FWO-Flanders), and ERC Grants QUTE (No. 647905) and ERQUAF (No. 715861). T.~Vieijra is supported as an `FWO-aspirant' under contract number FWO18/ASP/279.
\end{acknowledgements}

\FloatBarrier
\bibliography{SU2-RBM_arxiv.bib}
\onecolumngrid

\newpage
\revappendix

\section{Basis transformation}
To find eigenstates of the \SU~symmetry group with restricted Boltzmann machines (RBM), the spins in the system are coupled to well-defined angular momentum eigenstates $\ket{J\,M_J}$ of the total system. Two spins $\h{\mathbf{s}}_1$ and $\h{\mathbf{s}}_2$ can be coupled to a total angular momentum $\h{\mathbf{j}}_2=\h{\mathbf{s}}_1+\h{\mathbf{s}}_2$ via $3j$-symbols, or equivalently Clebsch-Gordan coefficients, as
\begin{equation}
\begin{split}
    \ket{s_1\,s_2;j_2\,m_{j_2}} & = 
     \sum_{m_{s_1}=-s_1}^{s_1}\sum_{m_{s_2}=-s_2}^{s_2} (-1)^{-s_1+s_2-m_{j_2}} \sqrt{2j_2+1} \tj{s_1}{s_2}{j_2}{m_{s_1}}{m_{s_2}}{-m_{j_2}} \ket{s_1\,m_{s_1},s_2\,m_{s_2}}\\
    & =\sum_{m_{s_1}=-s_1}^{s_1}\sum_{m_{s_2}=-s_2}^{s_2}\braket{s_1\,m_{s_1},s_2\,m_{s_2}|s_1\,s_2;j_2\,m_{j_2}} \ket{s_1\,m_{s_1},s_2\,m_{s_2}}.
    \end{split}
    \label{eq:2s_coup}
\end{equation}
Eq.~\eqref{eq:2s_coup} relates product states of two basis states $\ket{s_1\,m_{s_1}}$ and $\ket{s_2\,m_{s_2}}$ of the Hilbert spaces $\mathcal{H}_1$ and $\mathcal{H}_2$ to states $\ket{s_1\,s_2;j_2\,m_{j_2}}$ representing the tensor product space $\mathcal{H}_1 \otimes \mathcal{H}_2$. The states $\ket{s_1\,s_2;j_2\,m_{j_2}}$ are angular-momentum eigenstates, implying that $\h{\mathbf{j}}_2^2 \ket{s_1\,s_2;j_2\,m_{j_2}} = j_2(j_2+1) \ket{s_1\,s_2;j_2\,m_{j_2}}$ and $\h{j}_{z_2} \ket{s_1\,s_2;j_2\,m_{j_2}} = m_{j_2} \ket{s_1\,s_2;j_2\,m_{j_2}}$. The orthonormality of the states $\ket{s_1\,s_2;j_2\,m_{j_2}}$ in the tensor product space $\mathcal{H}_1 \otimes \mathcal{H}_2$ follows readily from its definition and the use of the completeness relations of the tensor product basis
\begin{equation}
    \wh{\mathbb{1}} = \sum_{m_{s_1}=-s_1}^{s_1}\sum_{m_{s_2}=-s_2}^{s_2} \ket{s_1\,m_{s_1},s_2\,m_{s_2}}\bra{s_1\,m_{s_1},s_2\,m_{s_2}}.
    \label{eq:clebsch}
\end{equation}
The completeness of the coupled basis follows from the completeness of the product states and the fact that the coupled basis has the same number of orthonormal elements.

A system comprising $N$ spins can be coupled to a total angular momentum $\wh{\mathbf{J}}= \h{\mathbf{s}}_1 + \h{\mathbf{s}}_2 + ... + \h{\mathbf{s}}_N$ by sequentially using the coupling rule of Eq.~\eqref{eq:2s_coup} until all the spins are coupled to a total angular momentum state $\ket{J\,M_J}$. The sequential coupling of spins can be performed using different schemes.  We adopt a scheme that consists of coupling the spins $\h{\mathbf{s}}_i$ in a linear fashion from left to right.  Specifically, we start with an ancillary angular momentum state $\ket{j_0\,m_{j_0}} = \ket{0\,0}$ to which we couple $\ket{s_1\,m_{s_1}}$, resulting in $\ket{j_1=s_1\,m_{j_1}=m_{s_1}}$. Next, $\ket{s_2\,m_{s_2}}$ is coupled to $\ket{j_1\,m_{j_1}}$ resulting in $j_2\in\{0,1\}$. This is repeated until the end of the chain is reached, where we couple $\ket{s_N\,m_{s_N}}$ to $\ket{j_{N-1}\,m_{j_{N-1}}}$, resulting in $\ket{j_{N}\equiv J\,m_{j_{N}}\equiv M_J}$. This coupling scheme yields the set of intermediate total angular momenta $\{j_1,j_2,...,j_{N-1}\}$ as degrees of freedom. According to the angular momentum addition rules, the possible configurations of these intermediate total angular momenta have to fulfill the triangle inequalities $|j_{i-1} - s_{i}| \leq j_i \leq |j_{i-1} + s_{i}|$.

With this coupling scheme, the full basis transformation can be written as
\begin{equation}
\begin{split}
    & \ket{s_1\,...\,s_N\,j_1\,...\,j_{N-1};J\,M_J} = \\
    & \qquad \sum_{\{m_{s_i}\}}\sum_{\{m_{j_i}\}} \left( \prod_{i=1}^N (-1)^{-j_{i-1}+s_i-m_{j_i}} \sqrt{2j_i+1}\tj{j_{i-1}}{s_i}{j_i}{m_{j_{i-1}}}{m_{s_i}}{-m_{j_i}}  \right)\delta_{j_{N}J} \delta_{m_{j_{N}}M_{J}} \ket{s_1\,m_{s_1},s_2\,m_{s_2},...,s_N\,m_{s_N}}.
    \end{split}
    \label{eq:2s_cou}
\end{equation}
The  expression  of  Eq.~\eqref{eq:2s_cou}  is  obtained  by  repeatedly  using  Eq.~\eqref{eq:2s_coup}  to  couple  the  intermediate angular  momenta. The procedure is outlined  in the  above-mentioned coupling  scheme  and is graphically depicted in Fig.~1(a) of the  main  text.

From now on, we denote $\ket{j_0\,...\,j_{N-1};J\,M_J} \equiv \ket{s_1\,...\,s_N\,j_0\,...\,j_{N-1};J\,M_J}$, as the spin degrees of freedom $s_1, ..., s_N$ are fixed.  The basis $\ket{j_0\,...\,j_{N-1};J\,M_J}$ forms an orthonormal basis, as can be seen from
\begin{equation}
\begin{split}
    & \braket{j'_0\,j'_1\,...\,j'_{N-1};J\,M_J|j_0\,j_1\,...\,j_{N-1};J\,M_J} \\
    & 
    \begin{split} \qquad = \sum_{\{m_{s_i}\}}\sum_{\{m_{j_i}\}}\sum_{\{m_{j'_i}\}} & \left( \prod_{i=1}^N (-1)^{-j_{i-1}+s_i-m_{j_i}} \sqrt{2j_i+1}\tj{j_{i-1}}{s_i}{j_i}{m_{j_{i-1}}}{m_{s_i}}{-m_{j_i}} \right) \\
    & \left( \prod_{i=1}^N (-1)^{-j'_{i-1}+s_i-m_{j'_i}} \sqrt{2j'_i+1}\tj{j'_{i-1}}{s_i}{j'_i}{m_{j'_{i-1}}}{m_{s_i}}{-m_{j'_i}} \right)
    \end{split}\\
    & \qquad = \delta_{j_1,j'_1} \delta_{j_2,j'_2} ... \delta_{j_{N-1},j'_{N-1}},
    \end{split}
    \label{eq:ortho}
\end{equation}
where we have used the orthogonality relation of $3j$-symbols
\begin{equation}
\sum_{m_{j_{i-1}}} \sum_{m_{s_{i}}} (2j_{i} + 1)\tj{j_{i-1}}{s_{i}}{j_{i}}{m_{j_{i-1}}}{m_{s_{i}}}{m_{j_{i}}} \tj{j_{i-1}}{s_{i}}{j'_{i}}{m_{j_{i-1}}}{m_{s_{i}}}{m_{j'_{i}}} = \delta_{j'_{i},j_{i}} \delta_{m_{j'_{i}},m_{j_{i}}},
\end{equation}
from left to right.  The completeness of the basis follows from the fact that, given the product states of two complete bases, the Clebsch-Gordan coefficients relate these states to a complete basis of the coupled system.  Using the Clebsch-Gordan coefficients consecutively thus retains the completeness when all subsystems are described by a complete basis.

\section{F-move}
To find the matrix elements of the antiferromagnetic Heisenberg Hamiltonian 
\begin{equation}
    \widehat{H} = \sum_{i=1}^{N-1} \hat{\mathbf{s}}_i \cdot \hat{\mathbf{s}}_{i+1},
\end{equation}
the basis is recoupled in such a way that the physical spins $\h{\mathbf{s}}_i$ and $\h{\mathbf{s}}_{i+1}$ that are subjected to the interaction $\h{\mathbf{s}}_i \cdot \h{\mathbf{s}}_{i+1}$ are coupled to a total angular momentum $\wh{\mathbf{S}}=\h{\mathbf{s}}_i+\h{\mathbf{s}}_{i+1}$. This can be done by using an F-move, which is a recoupling using $6j$-symbols. 

\begin{figure}[h!]
\includegraphics{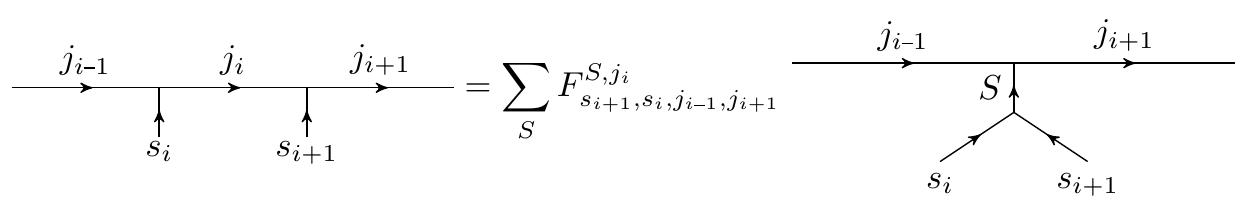}
    \caption{Depiction of the F-move which transforms between two orderings of the coupling of three angular momenta.} 
        \label{fig:Fmove}

\end{figure}
Graphically and mathematically, the F-move adopts the form of Fig.~\ref{fig:Fmove}, where 
\begin{equation}
    F^{S,j_{i}}_{s_{i+1},s_{i},j_{i-1},j_{i+1}} = (-1)^{s_i + s_{i+1} + j_{i-1} + j_{i+1}} \sqrt{(2S + 1)(2j_i + 1)}
    \begin{Bmatrix}
    s_{i+1} & s_{i} & S \\
    j_{i-1} & j_{i+1} & j_i 
    \end{Bmatrix}.
\end{equation}
The term between curly brackets is a $6j$-symbol. This yields the basis transformation
\begin{equation}
    \ket{j_0\,...\,j_{i-1}\,j_{i}\,j_{i+1}\,...\,j_{N-1};J\,M_J} = \sum_{S} F^{S,j_{i}}_{s_{i+1},s_{i},j_{i-1},j_{i+1}} \ket{j_0\,...\,j_{i-1}\,S\,j_{i+1}\,...\,j_{N-1};J\,M_J}.
    \label{eq:recoupling}
\end{equation}
With $\h{\mathbf{s}}_i \cdot \h{\mathbf{s}}_{i+1} = \frac{1}{2}(S^2 - s_i^2 - s_{i+1}^2)$ and the recoupling scheme of Eq.~\eqref{eq:recoupling} one finds
\begin{equation}
\begin{split}
    & \frac{1}{2}(S^2 - s_i^2 - s_{i+1}^2) \ket{j_0\,...\,j_{i-1}\,j_{i}\,j_{i+1}\,...\,j_{N-1};J\,M_J} \\
    & = \sum_{S} \frac{1}{2}(S(S+1) - s_i(s_i+1) - s_{i+1}(s_{i+1}+1))F^{S,j_{i}}_{s_{i+1},s_{i},j_{i-1},j_{i+1}} \ket{j_0\,...\,j_{i-1}\,S\,j_{i+1}\,...\,j_{N-1};J\,M_J}.
    \end{split}
    \label{eq:rec}
\end{equation}
Using the recoupling defined in Eq.~\eqref{eq:recoupling}, the matrix elements are found in the basis labeled by states $\ket{j_0\,j_1\,...\,j_{N-1};J\,M_J}$
\begin{equation}
    \begin{split}
    & \braket{j_0\,...\,j_{i-1}\,j'_i\,j_{i+1}\,...\,j_{N-1};J\,M_J|\h{\mathbf{s}}_i \cdot \h{\mathbf{s}}_{i+1}|j_0\,...\,j_{i-1}\,j_i\,j_{i+1}\,...\,j_{N-1};J\,M_J} \\
    & \qquad = \sum_{S} \frac{1}{2}(S(S+1) - s_i(s_i+1) - s_{i+1}(s_{i+1}+1))F^{S,j_{i}}_{s_{i+1},s_{i},j_{i-1},j_{i+1}} F^{S,j'_{i}}_{s_{i+1},s_{i},j_{i-1},j_{i+1}}.
    \end{split}
    \label{eq:melem}
\end{equation}
As can be seen in Eq.~\eqref{eq:melem}, the matrix elements of the operator $\h{\mathbf{s}}_i \cdot \h{\mathbf{s}}_{i+1}$ are independent of the values of $j_{k \notin \{i-1,i,i+1\}}$, are diagonal in $j_{k \in \{i-1,i+1\}}$ and are not diagonal in $j_i$.

\section{\SU-symmetry}
The \SU~symmetry group is a continuous symmetry group. For spin degrees of freedom, it is represented by the unitary operators 
\begin{equation}
    \wh{U}(\bm{\theta}) = e^{i(\theta_x \h{s}_x + \theta_y \h{s}_y + \theta_z \h{s}_z)} = e^{i\h{\mathbf{s}} \cdot \bm{\theta}},
\label{eq:U}
\end{equation}
where $\h{s}_{x,y,z}$ are the generators of \SU~and $\bm{\theta}$ is a normalised vector in three-dimensional space. For spin-\sfrac{1}{2}, $\h{s}_{x,y,z} = \h{\sigma}_{x,y,z}/2$, where $\h{\sigma}_{x,y,z}$ are the two-dimensional Pauli-matrices
\begin{equation}
\h{\sigma}_x = 
\begin{pmatrix}
0 & 1 \\
1 & 0 \\
\end{pmatrix}
,\quad
\h{\sigma}_y =
\begin{pmatrix}
0 & -i \\
i & 0 \\
\end{pmatrix}
,\quad
\h{\sigma}_z =
\begin{pmatrix}
1 & 0 \\
0 & -1 \\
\end{pmatrix}.
\label{eq:paulimatrix}
\end{equation}
For spin-$1$, $\h{s}_{x,y,z} $ are the three-dimensional matrices
\begin{equation}
\h{s}_x = 
\begin{pmatrix}
0 & \frac{1}{\sqrt{2}} & 0 \\
\frac{1}{\sqrt{2}} & 0 & \frac{1}{\sqrt{2}} \\
0 & \frac{1}{\sqrt{2}} & 0 \\
\end{pmatrix}
,\quad
\h{s}_y =
\begin{pmatrix}
0 & \frac{-i}{\sqrt{2}} & 0 \\
\frac{i}{\sqrt{2}} & 0 & \frac{-i}{\sqrt{2}} \\
0 & \frac{i}{\sqrt{2}} & 0 \\
\end{pmatrix}
,\quad
\h{s}_z =
\begin{pmatrix}
1 & 0 & 0 \\
0 & 0 & 0 \\
0 & 0 & -1 \\
\end{pmatrix}.
\label{eq:paulimatrix}
\end{equation}
The representation of the Hilbert space in terms of basis states with a total angular momentum forms an irreducible representation with respect to the \SU~symmetry, leaving the total angular momentum invariant and transforming the angular momentum projection degrees of freedom. For a recoupling of spins using Clebsch-Gordan coefficients
\begin{equation}
    \ket{s_1\,s_2;j\,m_j} = \sum_{m_{s_1},m_{s_2}}{\braket{s_1\,m_{s_1},s_2\,m_{s_2} | j\,m_j}\ket{s_1\,m_{s_1},s_2\,m_{s_2}}},
\end{equation} 
acting with the unitary operator $\wh{U}(\bm{\theta}) = e^{i\h{\mathbf{j}} \cdot \bm{\theta}} = e^{i(\h{\mathbf{s}}_1 + \h{\mathbf{s}}_2) \cdot \bm{\theta}}$ on both sides of the equation yields
\begin{equation}
    e^{i\h{\mathbf{j}} \cdot \bm{\theta}} \ket{s_1\,s_2;j\, m_j} = \sum_{m_{s_1},m_{s_2}}{\braket{s_1\,m_{s_1},s_2\,m_{s_2} | j\,m_j} e^{i(\h{\mathbf{s}}_1 + \h{\mathbf{s}}_2) \cdot \bm{\theta}} \ket{s_1\,m_{s_1},s_2\,m_{s_2}}}.
\end{equation}
This means that the coupling which transforms $\ket{s_1\,m_{s_1},s_2\,m_{s_2}}$ in $\ket{s_1\,s_2;j\, m_j}$ also transforms $e^{i(\h{\mathbf{s}}_1 + \h{\mathbf{s}}_2) \cdot \bm{\theta}} \ket{s_1\,m_{s_1},s_2\,m_{s_2}}$ in $e^{i\h{\mathbf{j}} \cdot \bm{\theta}} \ket{s_1\,s_2;j\, m_j}$.  Using this result sequentially on the coupling used in this work, we see that the coupling which transforms $\ket{s_1\,m_{s_1};...;s_N\,m_{s_N}}$ in $\ket{s_1\,...\,s_N\, j_0\,j_1\, ...\, j_{N-1};J\, m_J}$ also transforms $e^{i(\h{\mathbf{s}}_1 + ... + \h{\mathbf{s}}_N) \cdot \bm{\theta}} \ket{s_1\,m_{s_1};...;s_N\,m_{s_N}}$ in $e^{i\h{\mathbf{J}} \cdot \bm{\theta}} \ket{s_1\,...\,s_N\,j_1\,...\,j_{N-1};J\, m_J}$.  In the specific case where we couple to the angular momentum state $\ket{J=0,M_J=0}$, $e^{i\h{\mathbf{0}}\cdot \bm{\theta}} = \wh{\mathbb{1}}$, which does indeed leave the coupled state invariant.  This proves that the variational ground-state wave function is manifestly invariant under \SU~transformations since every basis state is itself invariant under \SU~transformations.
\section{Basis cut-off}
\begin{figure}[h]
\includegraphics{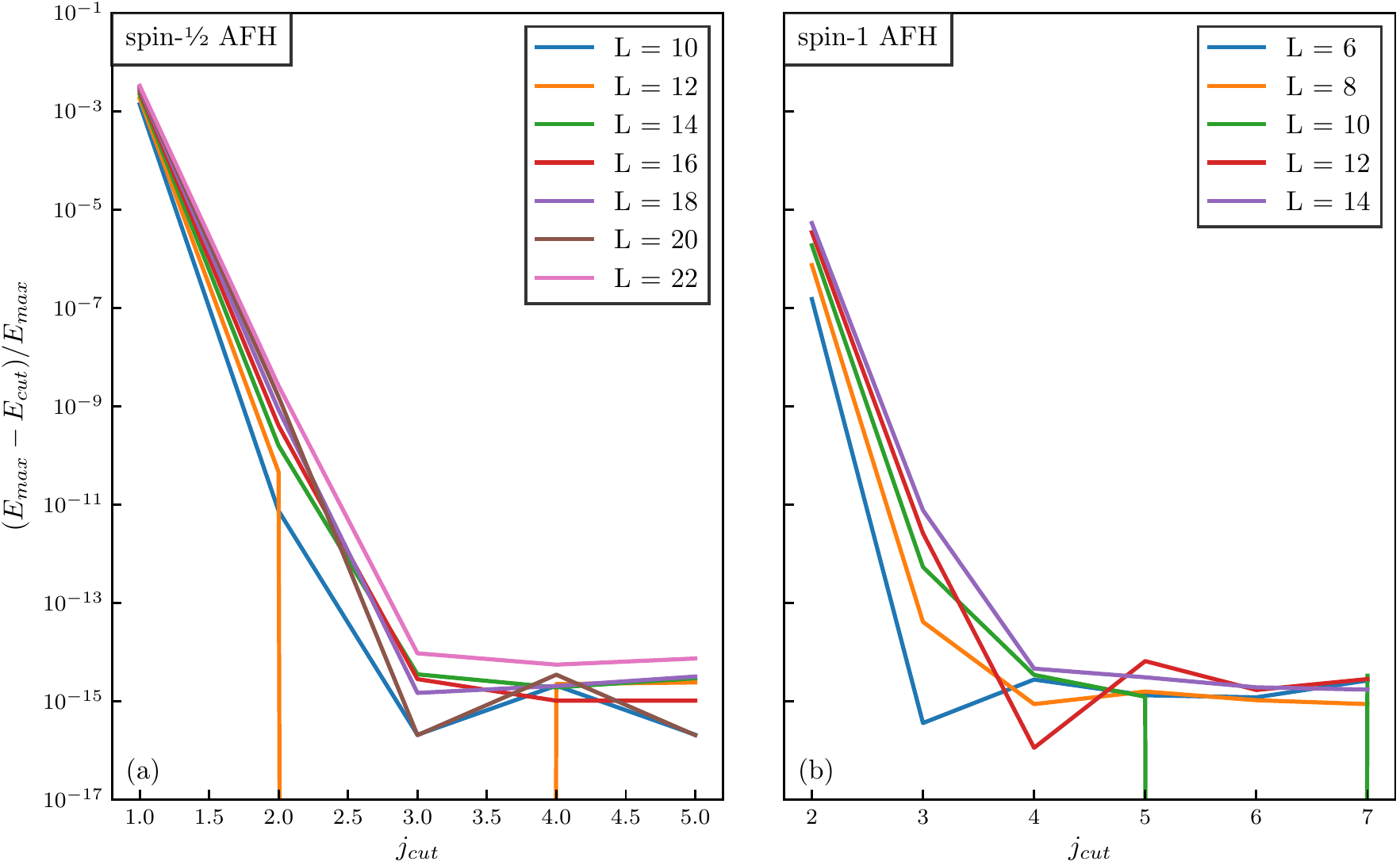}
  \caption{Exact diagonalization energy in the basis with intermediate angular momentum cut-off $E_{cut}$, relative to that of the full basis $E_{max}$, as a function of the cut-off $j_{cut}$.  (a) spin-$\sfrac{1}{2}$ AFH, (b) spin-$1$ AFH.  For all observed system sizes, $(E_{max} - E_{cut})/ E_{max}$ decreases rapidly and settles to machine precision for $j_{cut}=3$ (spin-$\sfrac{1}{2}$ AFH) and $j_{cut}=4$ (spin-$1$ AFH).}
    \label{fig:ED_cut}
\end{figure}
The basis transformation used in this paper introduces the intermediate angular momenta $j_{i\in\{1,...,N-1\}}$ as degrees of freedom.  In the case of spin-\sfrac{1}{2}, these can take on values $j_i \in \{0, \sfrac{1}{2}, 1, \sfrac{3}{2}, 2, ...\}$, while in the case of spin-$1$, they can take on values $j_i \in \{0,1,2,3,...\}$. As explained earlier, the intermediate degrees of freedom need to satisfy the triangle relations $\left| j_{i-1} - s_{i} \right| \leq j_i \leq \left| j_{i-1} + s_{i} \right|$.  These constraints induce a maximal value $j_{max}$ for the degrees of freedom, which is proportional with system size $N$.  Because the RBM parameters explicitly depend on the input values, the number of weights also increases linearly with $j_{max}$.\\

The eigenstate with minimal eigenvalue of the individual terms of the Hamiltonian is the state where the two spins involved couple to a singlet. Hence, one expects physically that the antiferromagnetic Heisenberg Hamiltonian favors the coupling of spins to a small total angular momentum. This motivates the introduction of a cut-off $j_{cut}$ for the maximal value of the intermediate angular momenta $j_i$. In Fig.~\ref{fig:ED_cut}, the effect of introducing the cut-off $j_{cut}$ is inspected by comparing the relative energy error of the ground state energy of the AFH with and without this constraint, for different system sizes.  The results are obtained with exact diagonalization.  We see that the relative energy error decreases rapidly and settles to machine precision at $j_{cut} = 3$ for the spin-\sfrac{1}{2} model and $j_{cut} = 4$ for the spin-1 model.  Moreover, the dependence on system size is negligible, even for the largest systems, indicating that the cut-off is inherent to the physics of the system.

\section{Wave function structure}
\begin{figure}[t]
\includegraphics{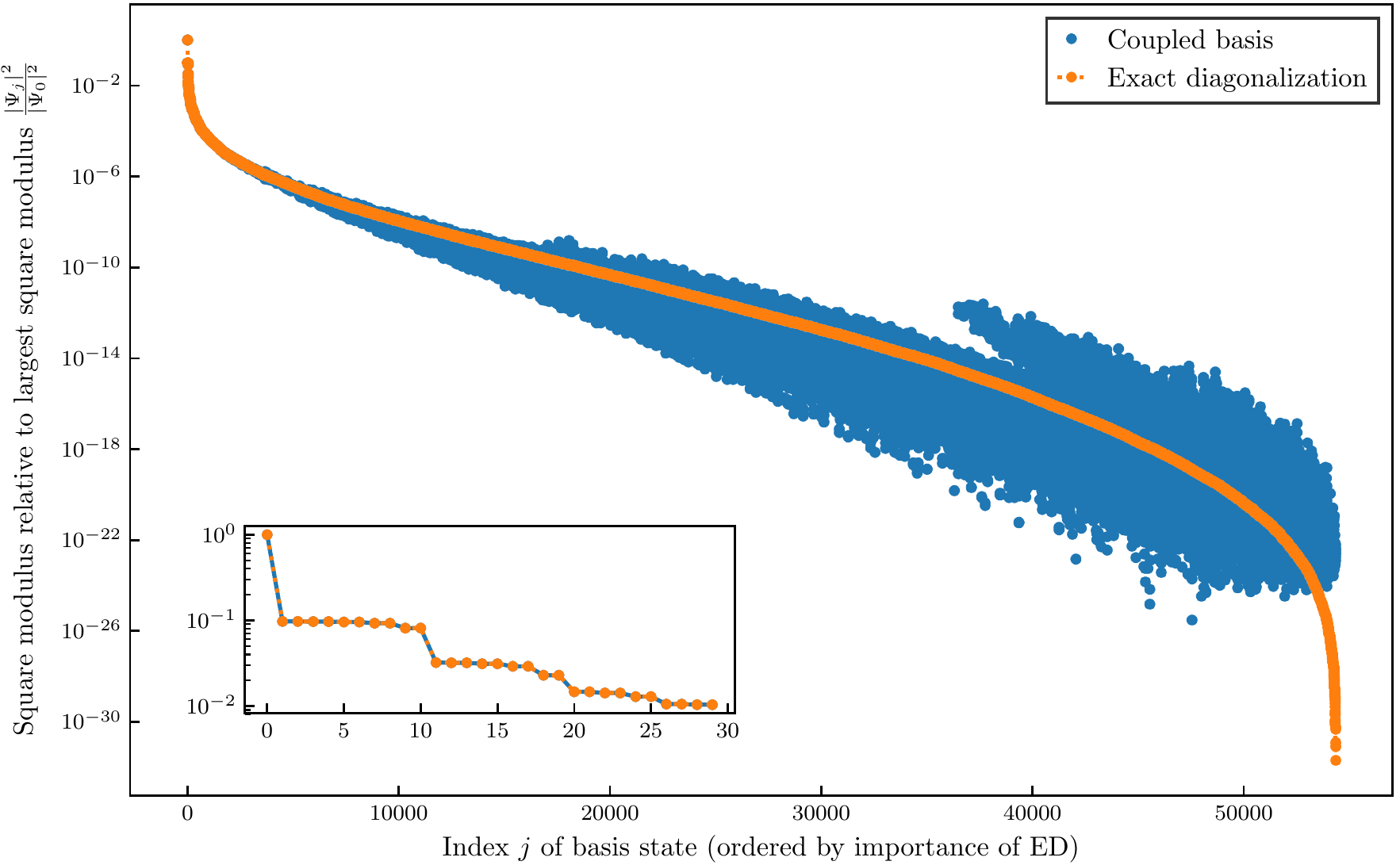}
  \caption{Square modulus relative to the largest square modulus of the basis states $|\Psi_j|^2/|\Psi_0|^2$ of the spin-\sfrac{1}{2} AFH.  The importance measure $|\Psi_j|^2/|\Psi_0|^2$ is plotted according to the index in the ordered list from high to low of the ED coefficients.  Both ED (orange line) and variational (blue line) coefficients are shown.  The inset shows a detail of the first $30$ square moduli.  Note the relation with the basis states shown in Fig.~\ref{fig:confs_both}.}
    \label{fig:importance}
\end{figure}
\begin{figure}[t]
\includegraphics{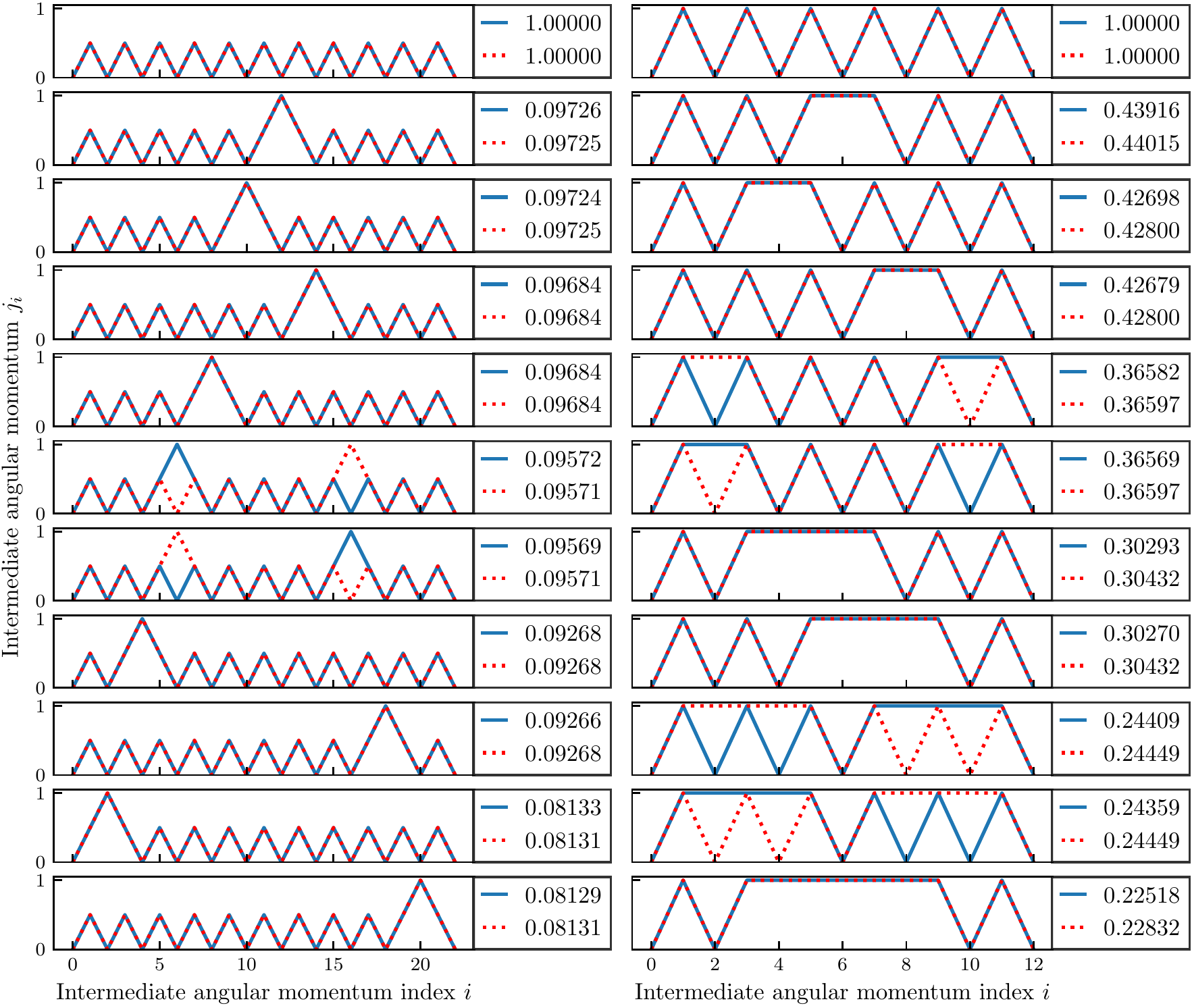}
\caption{The 11 most important configurations of the spin-$\sfrac{1}{2}$ (left column) and spin-1 (right column) AFH.  Importance is measured as the square modulus of the expansion coefficient of the configuration.  Red dashed lines are ED results, blue lines are results using the RBM as a variational ansatz.  The legend denotes the square modulus of the depicted expansion coefficient, divided by that of the most important one.}
\label{fig:confs_both}
\end{figure}

A central question in variational many-body state optimization is the level of accuracy of observables other than the energy (which is used as the optimization criterion). More information can be obtained by looking at how close the individual expansion coefficients of the chosen basis functions lie to the exact expansion coefficients. Hence, we compare the expansion coefficients of the RBM model to those of exact diagonalization in the coupled basis. An interesting follow-up question is which configurations of intermediate spins have the highest importance in the basis expansion of the ground state, where we define the importance as the modulus of the expansion coefficient. We denote $\Psi_j$ as the expansion coefficient with the $j$-th largest modulus. Because the RBM wave function is not normalized, we compare the square modulus of the expansion coefficients relative to that with the largest square modulus, i.e. $|\Psi_j|^2/|\Psi_0|^2$. Fig.~\ref{fig:importance} shows the relative importance of the expansion coefficients $|\Psi_j|^2/|\Psi_0|^2$ as a function of $j$ for both the variational wave function and the wave function obtained with exact diagonalization. For $|\Psi_j|^2/|\Psi_0|^2 \gtrsim 10^{-6}$ the expansion coefficients coincide, while for $|\Psi_j|^2/|\Psi_0|^2 \lesssim 10^{-6}$ they start to diverge slightly. The left column of Fig.~\ref{fig:confs_both} shows from top to bottom the $11$ most important configurations of the spin-\sfrac{1}{2} AFH in the coupled basis.  Using the recoupling of spins, we can provide an interpretation to the depicted configurations.  The most important configuration can be recoupled to the case where, from left to right, every two neighbouring spins are coupled to a singlet. Note that this is a particular case of the resonating valence bond state.  The next $10$ configurations can all be recoupled to a background consisting of singlets (as in the first configuration), but where two neighbouring singlets are excited to two neighbouring triplets, which couple together to a singlet. The results for the spin-1 AFH are shown in the right column of Fig.~\ref{fig:confs_both}, and are qualitatively similar to the spin-\sfrac{1}{2} case.

\section{Correlations}
\begin{figure}[t]
\includegraphics{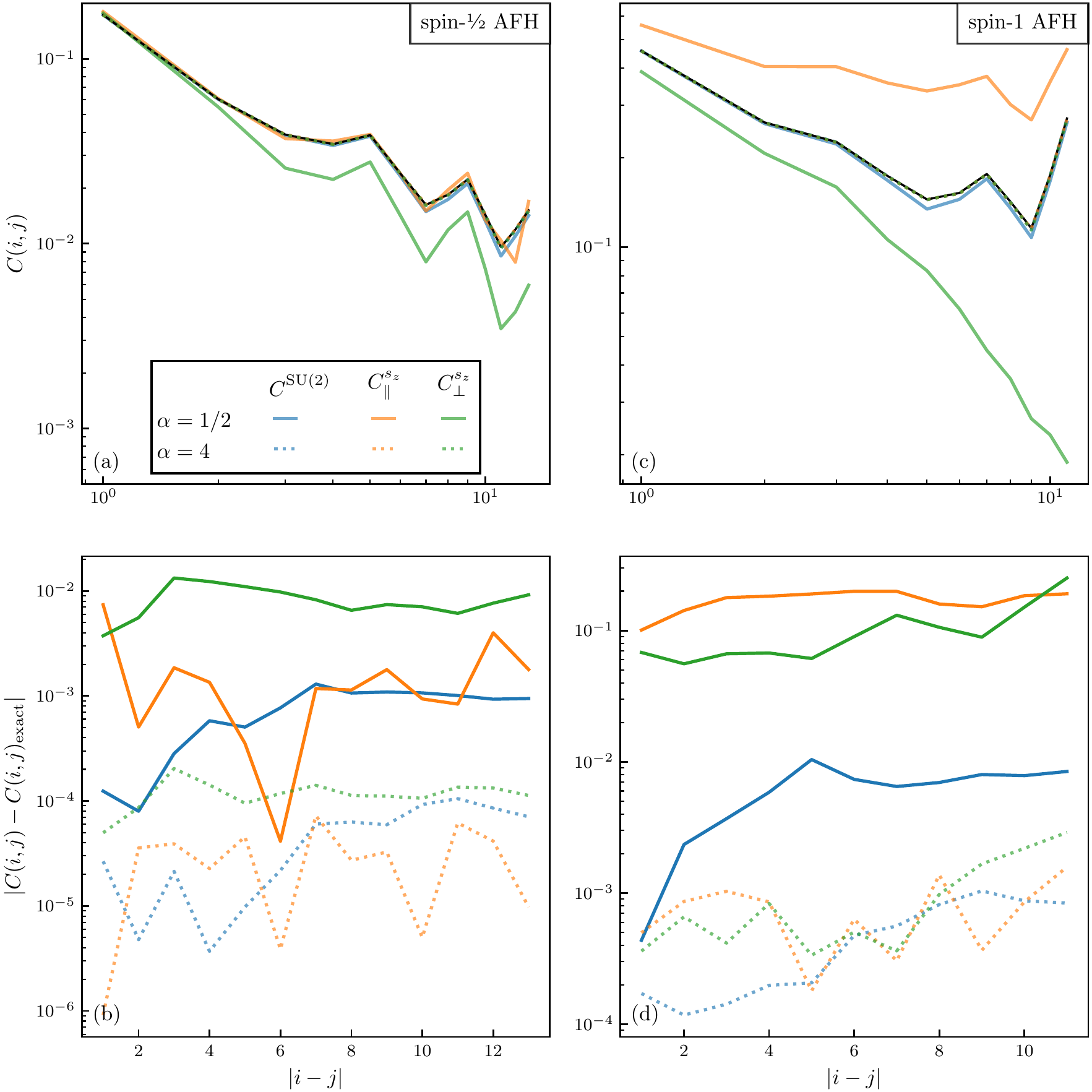}
  \caption{Correlation functions for the spin-$\sfrac{1}{2}$ and spin-$1$ Heisenberg model with low ($\alpha=0.5$) and high ($\alpha=4$) number of hidden units.  The wave functions are the same as those of Fig.~(2) in the main text.  For the $s_z$ basis, we show the longitudinal component $C^{s_z}_{\parallel}(i,j)= C^{s_z}_{z,z}(i,j)$ and the transverse component $C^{s_z}_{\perp}(i,j) = (C^{s_z}_{x,x}(i,j) + C^{s_z}_{y,y}(i,j))/2$ separately.  For the coupled basis, we show $C^{\textrm{SU(2)}}(i,j) = C^{\textrm{SU(2)}}_{x,x}(i,j) = C^{\textrm{SU(2)}}_{y,y}(i,j) = C^{\textrm{SU(2)}}_{z,z}(i,j)$. (a) Correlation function as a function of distance for the spin-$\sfrac{1}{2}$ Heisenberg model.  The black line represents the exact correlation function. (b) Absolute error relative to the exact values of the correlation function of the spin-$\sfrac{1}{2}$ Heisenberg model. (c) and (d) Same as (a) and (b) but for the spin-$1$ Heisenberg model.}
    \label{fig:correlations}
\end{figure}
To assess the advantage of imposing symmetries in RBM wave functions, we compute the spin-spin correlation functions for the wave functions discussed in the main text.  The spin-spin correlation functions are defined as
\begin{equation}
    C_{\beta, \gamma}(i,j) = \langle \hat{s}_i^{\beta} \hat{s}_j^{\gamma} \rangle,
\end{equation}
with $\beta,\gamma \in \{x,y,z\}$ the different directions of the spin operator $\hat{\mathbf{s}}$.  For the Heisenberg model, the components of the correlation function with $\beta=\gamma$ should be equal due to the inherent \SU-symmetry.  To prove that, in the coupled basis, this principle holds by construction, we make use of the Wigner-Eckart theorem
\begin{equation}
    \braket{J'\,M'_J|\tilde{T}_{l,m_l}|J\,M_J} = \braket{J'\,M'_J;l\,m_l|J\,m_J}\braket{J'||\tilde{T}_l||J}.
    \label{eq:wigner_eckart}
\end{equation}
The Wigner-Eckart theorem states that the overlap of a $\tilde{T}_{l,m_l}$ spherical tensor operator of rank $l$ and component $m_l$ between two angular momentum eigenstates $\ket{J\,m_J}$ and $\ket{J'\,m'_J}$ is equal to the Clebsch-Gordan coefficient $\braket{J'\,m'_J;l\,m_l|J\,m_J}$ multiplied with a reduced matrix element $\braket{J'||\widehat{T}_l||J}$, independent of the angular momentum projections.  As one can see from Eq.~\eqref{eq:wigner_eckart}, using states with $J=J'=0$, which is the case for the ground state in the coupled basis, all expectation values of spherical operators with $l \neq 0$ are identically zero.

We write the $9$ Cartesian components of the correlation functions in terms of the operators in the spherical basis via a linear transformation.  Here, we use the definitions of Ref.~\cite{emsly_nuclear_1985}.  In particular, the spherical operator with $l=0$ and $m_l=0$ is defined as
\begin{equation}
    \tilde{C}_{0,0}(i,j) = -\frac{1}{\sqrt{3}}\left( C_{x,x}(i,j) + C_{y,y}(i,j) + C_{z,z}(i,j) \right).
\end{equation}
The Cartesian operators with $\beta=\gamma$ are related to the spherical operators by
\begin{equation}
    \begin{split}
        &C_{x,x}(i,j) = -\frac{1}{\sqrt{3}}\left( \tilde{C}_{0,0}(i,j) - \frac{1}{\sqrt{6}}\tilde{C}_{2,0}(i,j) + \frac{1}{2} (\tilde{C}_{2,2}(i,j) + \tilde{C}_{2,-2}(i,j)) \right) \\
        &C_{y,y}(i,j) = -\frac{1}{\sqrt{3}}\left( \tilde{C}_{0,0}(i,j) - \frac{1}{\sqrt{6}}\tilde{C}_{2,0}(i,j) - \frac{1}{2} (\tilde{C}_{2,2}(i,j) + \tilde{C}_{2,-2}(i,j)) \right) \\
        &C_{z,z}(i,j) = -\frac{1}{\sqrt{3}}\left( \tilde{C}_{0,0}(i,j) + \sqrt{\frac{2}{3}}\tilde{C}_{2,0}(i,j) \right).
    \end{split}
    \label{eq:cartesian_ops}
\end{equation}
Using the Wigner-Eckart theorem for the Cartesian operators of Eq.~\eqref{eq:cartesian_ops}, we find that, as only the spherical component with $(l=0,m_l=0)$ contribute, all three Cartesian components in Eq.~\eqref{eq:cartesian_ops} are equal to a third of the scalar product $\mathbf{\hat{s}}_i \cdot \mathbf{\hat{s}_j}$.

In Fig.~\ref{fig:correlations}, we show the correlation functions and the absolute error of the correlation functions with respect to the exact values for both the spin-$\sfrac{1}{2}$ (Fig.~\ref{fig:correlations}~(a-b)) and spin-$1$ (Fig.~\ref{fig:correlations}~(c-d)) Heisenberg model, as described in the main text.  In particular, we show the correlation functions $C(i,j)$ as a function of distance $|i-j|$ for the lowest ($\alpha=0.5$) and for the highest ($\alpha=4$) number of hidden units.  For the $s_z$-basis, we show the longitudinal ($C^{s_z}_{\parallel}(i,j)= C^{s_z}_{z,z}(i,j)$) and transverse ($C^{s_z}_{\perp}(i,j) = (C^{s_z}_{x,x}(i,j) + C^{s_z}_{y,y}(i,j))/2$) correlation functions separately.  For the coupled basis, we show $C^{\textrm{SU(2)}}(i,j) = C^{\textrm{SU(2)}}_{x,x}(i,j) = C^{\textrm{SU(2)}}_{y,y}(i,j) = C^{\textrm{SU(2)}}_{z,z}(i,j)$.  For the correlations in the spin-$\sfrac{1}{2}$ Heisenberg model, one can notice that the RBM with $\alpha=0.5$ in the $s_z$-basis breaks the symmetry between the longitudinal and transverse components and is biased for the correlations in the longitudinal direction.  This is a direct consequence of using the $s_z$-basis as input states for the variational wavefunction.  For the RBM in the coupled basis, the correlations lie close to the exact values, even for $\alpha=0.5$.  For $\alpha=4$, both RBMs perform well, with absolute errors of $\mathcal{O}(10^{-5})$.  However, the inequality between the transverse and longitudinal components in the $s_z$-basis persists. For the correlations with $\alpha=0.5$ in the spin-$1$ Heisenberg model, the deviations are much larger in the $s_z$-basis, also for the longitudinal component.  While the scalar product of spins are reasonably well described, the individual components are largely over- or underestimated.  By construction, this is not the case in the coupled basis, which is further evidence for the power of using coupled irreducible representations of the symmetry groups of the model as basis states for the variational model.  

In all our experiments, the dependence of the error on the correlation functions on distance is weak.  This is in contrast with the correlation functions in the transverse field Ising model, which were recently studied in ref.\cite{collura2019descriptive}.  However, as the models and system sizes are different, it is difficult to compare both results.  

\section{Hyperparameters}
Several hyperparameters have to be chosen for the variational optimization of RBMs. These hyperparameters are the learning rate $\eta$ of the optimization algorithm (stochastic gradient descent), the diagonal shift $d$ of the stochastic reconfiguration algorithm, and the standard deviation $\sigma$ of the normal distribution (centered at 0) from which the parameters of the RBM are initialized.  In order to find the set of hyperparameters which enables the energy to reach its lowest values, we draw a number (order 100) of random combinations of hyperparameters, and use these to optimize RBM wave functions.  Typically, we sample $\eta$ uniformly in the range $\eta \in [0.01,0.2]$, $\log (d)$ uniformly in the range $\log (d) \in [-3,-1]$, and $\sigma$ uniformly in the range $\sigma \in [0.01,0.1]$.  We found that the energy after optimization is largely independent of $d$ and $\sigma$, but depends strongly on the learning rate $\eta$.  For the experiments in the coupled basis, we reached the lowest energies with $\eta \approx 0.1$, while for those in the $s_z$-basis, we used learning rates $\eta \in [0.01,0.1]$.  Finally, we used parallel tempering~\cite{swendsen_replica_1986} for the optimization of the spin-1 AFH.  The parallel tempering algorithm uses a set of Monte Carlo chains (replicas) that all sample the probability distribution raised to a power between $0$ and $1$. As a result, the chains with lower power sample a smoother distribution than those with higher power.  By mixing configurations between neighbouring chains (according to a Monte Carlo step), the chain with power $1$ (the physical distribution which is used to measure observables) samples a richer region of the phase space.  The number of replicas in the parallel tempering algorithm was 50 and were chosen with uniformly separated powers.
\FloatBarrier

\end{document}